\documentclass[flEquation,usenatbib]{mnras}

\usepackage{newtxtext,newtxmath}
\usepackage[T1]{fontenc}

\DeclareRobustCommand{\VAN}[3]{#2}
\let\VANthebibliography\thebibliography
\def\thebibliography{\DeclareRobustCommand{\VAN}[3]{##3}\VANthebibliography}

\usepackage{graphicx}	% Including figure files
\usepackage{xcolor,amsmath,listings,orcidlink,siunitx,epigraph}
\usepackage[flushmargin]{footmisc}

\definecolor{codegreen}{rgb}{0,0.6,0}
\definecolor{codegray}{rgb}{0.3,0.3,0.3}
\definecolor{codepurple}{rgb}{0.58,0,0.82}
\definecolor{backcolour}{rgb}{0.975,0.975,0.975}

\lstdefinestyle{mystyle}{
    backgroundcolor=\color{backcolour},   
    commentstyle=\color{codegreen},
    keywordstyle=\color{magenta},
    numberstyle=\tiny\color{codegray},
    stringstyle=\color{codepurple},
    basicstyle=\ttfamily\footnotesize,
    breaklines=true,                 
    captionpos=b,                    
    keepspaces=true,                 
    numbers=left,                    
    numbersep=2pt,                  
    tabsize=2
}
\title[Galactic Outflow Line Profiles]{Galactic Outflow Emission Line Profiles:\\Evidence for Dusty, Radiatively-Driven Ionized Winds in Mrk 462}

\author[S. R. Flury et al.]{
Sophia R. Flury\orcidlink{0000-0002-0159-2613},$^{1}$\thanks{E-mail: sflury@umass.edu}
Edward C. Moran,$^{2}$
Miriam Eleazer$^{1,2}$
\\
$^{1}$Department of Astronomy, University of Massechusetts Amherst, Amherst, MA 01002, United States\\
$^{2}$Astronomy Department, Wesleyan University, Middletown, CT 06459, United States
}

\date{Accepted XXX. Received YYY; in original form ZZZ}

\pubyear{2022}

\begin{document}
%\graphicspath{{./}{./plots/}}
\label{firstpage}
\pagerange{\pageref{firstpage}--\pageref{lastpage}}
\maketitle

% Abstract of the paper
\begin{abstract}
Over the past half century, gas outflows and winds have been observed as asymmetric emission lines in a wide range of astrophysical contexts, including galaxies and early-type stars. While P Cygni lines are modeled and understood with physically-motivated profiles under the Sobolev approximation, asymmetric nebular lines are not. Previous studies of galactic outflows using nebular emission lines have made physically unjustified assumptions about the shape of the line profile. These approaches limit assessment of outflow properties and do not connect observations to the underlying physics. The physical complexity of galactic outflows requires a more robust approach. In response to this need, we present a novel profile for modeling nebular emission lines which is generalized yet physically motivated and provides insight into the underlying mechanisms of galactic outflows. To demonstrate the usefulness of this profile, we fit it to the asymmetric nebular lines observed in the nuclear region of Mrk 462, a starburst-AGN composite galaxy. From analysis of the best-fit profile, we conclude that the observed profile arises from a dusty radiation-pressure-driven outflow with a terminal velocity of $750\rm~km~s^{-1}$. This outflow, while weak by some standards, is still sufficiently strong to regulate star formation and black hole growth in the host galaxy by removing gas from the inner few kiloparsecs. Outflows like the one we observe and characterize in Mrk 462 are crucial to our understanding of episodic gas-fueled activity in galactic nuclei, which undoubtedly plays a pivotal role in galaxy evolution.
\end{abstract}

% Select between one and six entries from the list of approved keywords.
% Don't make up new ones.

%%%%%%%%%%%%%%%%%%%%%%%%%%%%%%%%%%%%%%%%%%%%%%%%%%

%%%%%%%%%%%%%%%%% BODY OF PAPER %%%%%%%%%%%%%%%%%%
%\setlength{\epigraphwidth}{0.575\columnwidth}
%\epigraph{Wild Spirit, which art moving everywhere;\\Destroyer and preserver}{``Ode to the West Wind''\\Percy Shelley}

\section{Introduction}\label{sec:intro}

Gas outflows from galaxies are nearly ubiquitous and likely play a fundamental role in galaxy evolution from triggering bursts of star formation to temporary or even permanent quenching. So significant is the impact that the 2020 Decadal Survey identified the origins, nature, and effects of these outflows as one of the pressing open science questions in extragalactic astronomy. Understanding the underlying mechanisms and properties of outflows is important for establishing in detail the relationship between these energetic phenomena and their host galaxies.

%These outflows can affect galaxies via various feedback mechanisms (stellar winds, SNe, AGN disk winds...)

One of the most insightful diagnostics of outflow properties has been UV absorption-line spectroscopy. Many studies have simplified their assessment of these features using non-parametric approaches \citep[e.g.,][]{2010ApJ...717..289S,2011ApJ...730....5H,2015ApJ...809..147H,Henry_2015,2017A&A...605A..67C,2017ApJ...851L...9J} or by fitting one or more Gaussian or Voigt components to the line \citep[e.g.,][]{2002ApJ...570..588R,2005ApJ...621..227M,2005ApJ...631L..37R,2015ApJ...811..149C,2018A&A...616A..29G,2022ApJ...933..222X}. With detailed, physical models of outflows in recent years, absorption and P-Cygni line analysis (predominantly in the UV, although optical \ion{Na}{i} lines have been employed as well) has led to characterization of the mechanisms underlying the outflow \citep[][]{2001A&A...369..574V,2010ApJ...717..289S,2011ApJ...734...24P,2015ApJ...801...43S,2016MNRAS.463..541C,2018ApJ...860..143C,2021MNRAS.504.2051V}. However, with the anticipated loss of \emph{HST} (our sole UV observing facility) in the near future and accompanied by the advent of \emph{JWST}, emission lines in other rest-frame regimes must be used for detailed investigations of outflows going forward.

Most previous studies of outflows using nebular emission lines have provided little physical justification for the particular line profiles adopted. \citet{1980A&A....81..172P} and \citet{1981MNRAS.195..787P} established the use of multiple Gaussian components for achieving a better description of the line profile \citep[now commonplace in the galactic outflow literature, e.g.,][]{2011MNRAS.418.2032V,2012ApJ...754L..22A,2013MNRAS.436.2576L,2014MNRAS.441.3306H,2019MNRAS.485.3409G,2022arXiv220702245E,2022arXiv220913125A,2022arXiv221011575M}. 
%Villar-Mart\'in+ 2011, Davis+ 2012, Liu+ 2013, Harrison+ 2014, Gallagher+ 2018, Riffel+ 2021, Eggen+ 2022, Rodríguez+ in prep 
Others use the less physically-motivated Gauss-Hermite formalism from \citet{1993ApJ...407..525V} to merely ``account'' for line asymmetries instead of inferring properties of the outflow \citep[also prevalent in the literature, e.g.,][]{Salviander_2007,2010Ap&SS.327..239R,2012ApJ...756...51L,2019MNRAS.485.2054S,2021MNRAS.507...74R}. Alternatively, emission-line profile flux quantiles and line widths can be used characterize outflows \citep[e.g.,][]{1981ApJ...247..403H,1985MNRAS.213....1W,1991ApJS...75..383V,2013MNRAS.436.2576L,2014MNRAS.441.3306H,2020A&ARv..28....2V,2021MNRAS.507...74R}. While this non-parametric approach avoids the need for assumptions about the profile, its pure empiricism cannot fully connect observations with the underlying physics of the presumed outflow that produces them.

Seeing the lack of clear physical motivation, \citet{2017MNRAS.471.4061K} developed model profiles for atomic and molecular nebular lines that relate directly to the physics of outflows. However, their profiles require assumptions about the driving mechanisms (winds vs radiation), physics (point mass vs isothermal potentials) and geometry (conserved angle vs conserved solid angle) a priori. Despite the successful application of the \citet{2017MNRAS.471.4061K} profiles to the outflows observed in starburst galaxy M82, this highly tailored approach lacks the flexibility and wide applicability offered by the other methods described above. Moreover, it cannot provide general insight regarding the origin of the observed outflow.

There is obvious need for a model that properly relates the physics of galactic outflows to observed nebular emission lines \emph{without} making assumptions about the mechanisms beforehand. Further exacerbating the problem is the need for such a model to contain few parameters and to be flexible, i.e., to readily fit data without substantial tweaking. To address these needs, we have developed a novel model emission-line profile for use in analyzing ionized outflows. In \S\ref{sec:profmod}, we outline the formalism for a nebular emission-line profile derived from solutions to the equation of motion for interstellar winds. Then, in \S\ref{sec:proffit}, we demonstrate the application of this model to medium resolution rest-frame optical spectra of the nearby dwarf galaxy Mrk 462, which displays asymmetric emission-line profiles indicative of outflows (Eleazer et al. in prep). Interestingly, Mrk 462 possesses both an AGN and nuclear star formation \citep[][Eleazer et al. in prep]{2014AJ....148..136M}; it thus provides a unique context for the investigation of outflows in which the outflow origins and mechanisms are ambiguous (Eleazer et al. in prep). Finally, we assess the underlying physics implied by fits to the data in \S\ref{sec:disc}. From the best-fit velocity profile, we find that dusty radiation pressure is the most likely driver of the observed outflow, favoring a clumpy cloud geometry over a homogeneous one. We conclude that, while weak in relative terms, the outflow in Mrk 462 is sufficiently strong to expel gas from the galactic gravitational potential, thereby influencing star formation and/or AGN activity.

\section{A Nebular Line Profile for Galactic Outflows}\label{sec:profmod}

Many previous studies of photoionized outflows with nebular emission lines have demonstrated asymmetry in the line profile, notably in [\ion{O}{iii}]$\lambda5007$, where the blue wing is more prominent than the red \citep[e.g.,][although see \citealt{2021MNRAS.507...74R} for red-winged profiles]{1991ApJS...75..383V,2011MNRAS.418.2032V,2013MNRAS.436.2576L,2014MNRAS.441.3306H,2019MNRAS.485.3409G,2019ApJ...886...11L,2022arXiv220913125A}. %,2022ApJ...933..110X}
This asymmetric blue wing is ubiquitously attributed to outflows since neither rotation nor mergers can account for the observed profile \citep[e.g.,][]{1981ApJ...247..403H,2013MNRAS.436.2576L,2013ApJ...768...75R,2014MNRAS.441.3306H}. These observed asymmetric emission lines require a new, physically motivated line profile to obtain insight into the properties and driving mechanism(s) of the underlying outflow. As discussed in \S\ref{sec:intro}, such a line profile must satisfy two requirements. First, it must not presume the underlying physics. Second, it must be flexible by having clear formalism and few parameters to ensure versatility in the analysis of line profiles in a wide variety of systems.

Models for outflow profiles have been previously presented in the literature. However, these rely on assumptions about the underlying physics and geometry of the outflow, critically, the mechanism driving the outflow and whether energy or momentum is conserved in the process. Different mechanisms invoked in line profiles include layered turbulence \citep[][]{2009A&A...500..817B}, radiation pressure and hot winds \citep{2017MNRAS.471.4061K}. %Since we do not know a priori what produces the observed line profiles in Mrk 462, we must avoid assumptions implicit in adopting a physical model which presumes the underlying mechanisms. Given the ambiguous BPT classification of the integrated nebular line flux ratios in Mrk 462 \citep[see][Eleazer et al. in prep]{2014AJ....148..136M}, 
Our aim, motivated by the composite activity in Mrk 462, is to model outflows without making assumptions about how the observed line profiles are produced.
In objects like Mrk 462, the relationship between the observed line profile and the origins of the outflow becomes increasingly unclear: stellar feedback, AGN feedback, supernovae, or some combination of all three, could provide the initial acceleration. As a result, we require a physically-motivated line profile model which is wholly agnostic to the underlying mechanisms.

Below, we describe a velocity profile which is both a solution to the equation of motion for a galactic outflow and independent of the underlying physics. We also define a density profile to express the distribution of particles within the outflow. Then, we illustrate how to compute a nebular emission line profile from the velocity and density profiles. Finally, we elaborate on numerical methods to solve the relevant equations rapidly and accurately for the purpose of fitting our model to the observed emission lines.

\subsection{Velocity and Density Profiles\label{sec:profiles}}

Velocity profiles for winds can take on a variety of shapes arising from different treatments of the wind equations of motion \citep[e.g.,][]{1975ApJ...195..157C,1976ApJS...32..715L,1977PhDT.........3A,1977ApJ...213..737B}. The simple $\beta$ form of the velocity profile arises from a generalization of these solutions without loss of physical insight \citep{1979ApJS...39..481C,1986A&A...164...86P} and has become ubiquitous in the study of stellar winds and galactic outflows \citep[e.g.,][see similar formalism in \citealt{2010ApJ...717..289S,2011ApJ...734...24P}]{2001A&A...369..574V,2021MNRAS.504.2051V,2016MNRAS.463..541C,2017A&A...605A..67C,2018MNRAS.474.1688C}. It is thus justifiable to adopt
\begin{equation}\label{Equation:velprof}
    v(r) = v_\infty \left(1-\frac{R_0}{r}\right)^\beta
\end{equation}
where $R_0$ is the radial distance from the progenitor at which the outflow begins (the ``launch" radius) and $v_\infty$ is the asymptotic terminal velcity reached in the limit of $r\to\infty$. Typical values for $\beta$ range from $0.5$ \citep{1975ApJ...195..157C,1977PhDT.........3A,2016MNRAS.463..541C} to $4$ \citep{1977ApJ...213..737B} with $1$ often chosen for theoretical models of lines profiles produced by winds \citep[][]{1979ApJS...39..481C,1987ApJ...314..726L,2001A&A...369..574V} as values of 0.8-1 are consistent with observed stellar P Cygni profiles \citep[][]{1986A&A...164...86P}. Under the Sobolev approximation \citep[][although  \citealt{1944SvA.....21..143S,1957SvA.....1..678S} provide earlier renditions]{1960mes..book.....S}, the outflow velocity is solely responsible for producing the observed line profile because the velocity gradient is much steeper than any local velocities (e.g., thermal or turbulent Doppler broadening). Thus, the above velocity law can be taken as characteristic of the observed line profile \citep[cf.][for the same assumption]{1974ApJ...193..651N,1975ApJ...202..465M,1987ApJ...314..726L,2011ApJ...734...24P,2015ApJ...801...43S}.

For simplicity, gas number density is also assumed to follow a power law such that
\begin{equation}\label{Equation:denprof}
    n(r) = n_0 \left(\frac{R_0}{r}\right)^\alpha,
\end{equation}
where $n_0$ is the gas number density at $R_0$, again a common choice in the literature \citep[e.g.,][]{2015ApJ...801...43S,2016MNRAS.463..541C}. Simple algebra gives the density in the wind as a function of velocity such that
\begin{equation}
    n(v) = n_0 \left[1-\left(\frac{v}{v_\infty}\right)^{1/\beta}\right]^\alpha
\end{equation}
\citep[see Equation 11 in][for a comparable result]{2016MNRAS.463..541C}. As expressed here, typical values for $\alpha$ range from 1 to 9 with 2 being characteristic of a mass-conserving, isothermal sphere \citep[see, e.g.,][]{2005ApJS..160..115R,2011ApJ...734...24P,2016MNRAS.463..541C,2017MNRAS.471.4061K} and larger values indicating mass removal by mechanisms such as fountains \citep[e.g.,][]{2015ApJ...814...83L,2016MNRAS.463..541C}.

In some cases, continuity has been invoked to relate $\alpha$ and $\beta$, thereby reducing the number of free parameters \citep[e.g.,][]{2015ApJ...801...43S,2018ApJ...860..143C}. While it would be lucrative to reduce the dimensionality\footnote{Invoking continuity or any other physical assumption is not lucrative in the least for our purposes: we want to assume as little as possible a priori about the underlying physics.}, the combined profiles in Equations \ref{Equation:velprof}-\ref{Equation:denprof} prohibit doing so: the extremes of $r=R_0$ and $r\to\infty$ yield momentum densities $nv$ of zero while interior to these extremes, the momentum density is non-zero, a clear violation of continuity. As an alternative to continuity, we could simply assume some value for $\alpha$. However, leaving $\alpha$ free adds just one parameter more than either adding an additional Gaussian component or employing the Gauss-Hermite formalism. Moreover, leaving $\alpha$ free allows us to avoid any additional assumptions about the underlying physics of the outflow. We demonstrate below that $\alpha$ correlates with $v_\infty$ rather strongly (see \S\ref{sec:proffit}, Figure \ref{fig:corner}). Any presumed value or relationship for $\alpha$ thus biases the terminal outflow velocity obtained from the model. Therefore, we choose to include $\alpha$ as a distinct, free parameter in our model.

%We do find that the value of $\alpha$ has a relatively weak effect on the shape of the emergent line profile \citep[cf. similar results in][for absorption lines]{2016MNRAS.463..541C,2018MNRAS.474.1688C}, so one might argue that maintaining our agnosticism in the case of $\alpha$ is moot.

\subsection{Radiative Transfer\label{sec:radtrans}}

Gas at any given velocity will produce line emission by the same mechanism. The nebulae surrounding star-forming regions or AGN are optically thin to non-resonant line emission, so no scattering or escape effects need to be included when calculating the emergent flux density\footnote{Resonant lines, which are not optically thin, do require such treatment, typically under limiting assumptions to obtain an additional factor of $(1-\exp(-\tau))\tau^{-1}$ in the velocity integral \citep[e.g.,][]{2015ApJ...801...43S,2018ApJ...860..143C}.}. Adapting Equation 5.41 in \citet[][see also Equation 3 in \citealt{Peimbert_2017}]{source:osterbrock2006}, the total flux density at any wavelength $\lambda$ is given by the integral over all line-emitting clouds with a velocity $v$, projected velocity $u = (\lambda-\lambda_0)c/\lambda_0$, and emission coefficient $j_\lambda $ such that
\begin{equation}
    F_\lambda = 4\pi\int\limits_v j_\lambda(v) {\rm d}v %\exp\left[-\tau_\lambda(s)\right]
\end{equation}
for an optically thin emission line. Following the formalism for nebular emission lines in \citet{source:osterbrock2006}, $j_\lambda$ for nebular lines like H$\alpha$ or [\ion{O}{iii}] is given by 
\begin{equation}\label{Equation:emiscoef}
    j_\lambda = \varepsilon n_e n_{\chi}
\end{equation}
for a radiative electronic transition of energy $h\nu$, emissivity $\varepsilon$ related to the recombination rate or the collisional excitation rate \citep[see Equations 4.14 and 3.29, respectively, in][cf. Equation 3 in \citealt{Peimbert_2017}]{source:osterbrock2006}, electron density $n_e$ and density of ion $\chi$. Under the Sobolev approximation, local variations in $n_e$ relative to $n_{\chi}$ within a single cloud are negligible. Therefore, we assume that $n_e$ and $n_{\chi}$ are equivalent and can therefore be expressed by the density term $n(r)\propto n_e \propto n_{\chi}$. This assumption allows us to rewrite the emergent flux density as
\begin{equation}
    F_\lambda = F \phi_{\lambda}(\lambda_0,\alpha,\beta,v_\infty)
\end{equation}
where $\phi_\lambda$ is the line profile at any given $\lambda$ due to a combination of velocities and densities projected onto the line of sight, and $F=4\pi h\nu \varepsilon n_e n_{\chi}$ is the integrated flux of the outflow line emission.

To obtain the line profile, we consider a spherical geometry with density defined in terms of the radial distance from the outflow source. Doing so allows us to integrate the squared density profile over the range of all velocities $v/v_\infty=w$ that are projected to the observed velocity $u$. The maximum value of $w$ at any given $u$ in the line profile is necessarily unity since the outflow, by definition, cannot exceed its own terminal velocity. Because minimum value of $w$ is not a constant, we simply denote it as $w_\ell$. The resulting expression for $\phi$ is then
\begin{equation}\label{Equation:prof}
    \phi_\lambda = K^{-1} \int\limits_{w_{\ell}}^{1} \left[1-w^{1/\beta}\right]^{2\alpha} {\rm d}w
\end{equation}
where $K$ is a normalization constant ensuring $\int_\lambda\phi_\lambda{\rm d}\lambda = 1$.

%The minimum velocity $w_\ell$ of line-emitting gas contributing to the flux at $u$ cannot subceed the line-of-sight velocity $u$. However, occultation of the receding (blue-shifted) component of the outflow near the launch radius (i.e., the outflow source) can increase the minimum $w_\ell$ \citep[e.g.][]]{2015ApJ...801...43S,2018ApJ...860..143C}, which we discuss below.

For the blueshifted emission (i.e., outflow moving toward the observer), the minimum velocity $w_\ell$ is simply the absolute value of the observed velocity, giving
\begin{equation}
    w_{\ell} = |u|.
\end{equation}
For the redshifted emission (i.e., outflow receding away from the observer), occultation of the outflow by the outflow source (e.g., an AGN or young stellar population embedded in optically thick material) increases $w_\ell$ by a cosine projection of the source radius at the observed velocity $u$ \citep[cf.][for similar treatment with resonant lines]{2015ApJ...801...43S,2018ApJ...860..143C}. The resulting expression for $w_\ell$ is
\begin{equation}\label{Equation:wmin}
|u| = \frac{w_\ell}{x(w_\ell)}\sqrt{x(w_\ell)^2-1}
\end{equation}
where
\begin{equation}
    x = \frac{r}{R_0} = \left(1-w^{1/\beta}\right)^{-1}
\end{equation}
is the dimensionless radius.

Finally, the line profile normalization constant $K$ is given by integrating $\phi_\lambda$ over all possible $\lambda$ which satisfy $u\in[-1,1]$, i.e., $\lambda\in[\lambda_{-\infty},\lambda_{+\infty}]$ where  $\lambda_{\pm\infty} = \lambda_0 (1\pm v_\infty/c)$. Furthermore, the integral of $\phi_\lambda$ over $\lambda$ must be broken down into blueshifted ($\lambda\in[\lambda_{-\infty},\lambda_{0}]$) and redshifted ($\lambda\in[\lambda_{0},\lambda_{+\infty}]$) components to account for the difference in velocity limits (recall that this difference arises from occultation of the outflow by the outflow source).

\begin{figure}
    \centering
    \includegraphics[width=\columnwidth]{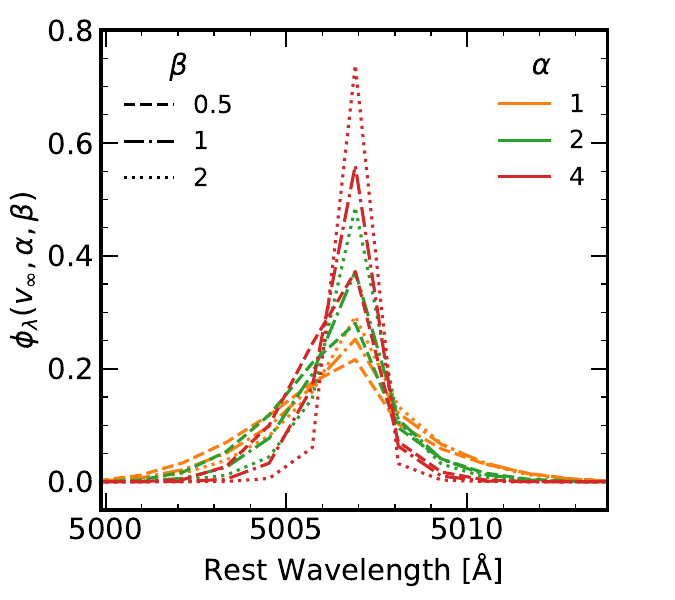}
    \caption{Effects of varying density profile slope $\alpha$ and velocity profile slope $\beta$ on the [\ion{O}{iii}] emission line profile emergent from the outflow. Example profiles assume a spectral resolution of $R=1688$ to emulate an observation with the \emph{Gemini}/GMOS B600+G5307 grating.}
    \label{fig:profexamp}
\end{figure}

As a reference to the reader, we illustrate effects of the density and velocity profile slopes on the emergent profile for the [\ion{O}{iii}]$\lambda5007$ emission line in Figure \ref{fig:profexamp}. Decreasing either the velocity profile slope $\beta$ or the density profile slope $\alpha$ increases the prominence of the profile wings at fixed $v_\infty$.

\subsection{Implementation\label{sec:implem}}

To obtain $w_\ell$ for the redshifted profile component ($\lambda\in[\lambda_0,\lambda_\infty]$), we use Brent's method with brackets of $w_\ell\in[u,1)$ to determine the value of $w_\ell$ which best satisfies Equation \ref{Equation:wmin}. Typically, Brent's method converges on a solution for $w_\ell$ within $\la10$ iterations for any given $u$. %The Newton-Raphson method provides similar solutions for $w_\ell$ when $\beta\in[0.4,1.6]$ but otherwise does not always converge on a solution.

To solve the integral term in Equation \ref{Equation:prof} for given $\lambda_0$, $\alpha$, $\beta$, and $v_\infty$, we use the Romberg integration method of adaptive step sizes for definite integrals. This approach optimizes the values of $w$ where the integrand is evaluated over $w\in[w_\ell,1]$ while minimizing the number of calculations necessary to converge on the solution. For $\phi_\lambda$, this method converges within $N\la8$ iterations with the total calculations of the integrand equal to $\sum_{i=0}^{N-1}2^i + 1$.

For the normalization coefficient $K$, we use a simple trapezoid integral of the $\phi_\lambda$ computed above. The trapezoid method is preferred over more sophisticated methods due to the cusp at $\lambda=\lambda_0$. The simplicity of the trapezoid method also requires very few calculations without significant loss of accuracy.

To facilitate reproducible results, we include in Appendix \ref{pdx:code} an example implementation of the emergent line profile formalism in python based on the numerical methods outline here.

\section{Application to Mrk 462}\label{sec:proffit}

Mrk 462 is a line-emitting dwarf galaxy with $M_\star=5.5\times10^9$ M$_\odot$ at a distance of $41.3$ Mpc. Both AGN and star-formation activity are evident in the rest-frame optical SDSS spectrum, with AGN signatures indicating the presence an intermediate mass black hole \citep{2014AJ....148..136M}. Eleazer et al (in prep) observed Mrk 462 for 1.4 hr on 2012/04/03 using \emph{Gemini}/GMOS with the B600+G5307 grating ($R=1688$ covering 4200-7100 \AA) to obtain integral field unit spectra covering the central kiloparsec of Mrk 462 in a $5^{\prime\prime}\times3.5^{\prime\prime}$ (1 kpc $\times$ 0.6 kpc at a distance of 41.3 Mpc) field of view. To maximize S/N, we proceed using the co-addition of all the \emph{Gemini}/GMOS IFU spaxels. For details of the data reduction and results, we refer the reader to their paper. During their efforts to determine the emission line fluxes, they discovered an asymmetric nebular emission line profile present in the strong forbidden narrow lines. The persistence of the asymmetric profile across the central kiloparsec of Mrk 462 indicates a large-scale outflow akin to those observed in AGN and starburst galaxies \citep[e.g.,][]{1981ApJ...247..403H,1990ApJS...74..833H,2011ApJ...729L..27R,2013ApJ...768...75R}. Mrk 462 thus presents a unique opportunity to demonstrate the use of our model to derive outflow properties from nebular emission line profiles. %While our initial motivation arises from a desire to understand the outflows we observe in Mrk 462, our ultimate goal is to demonstrate our methods for future studies of galactic outflows in a far broader context.

To begin our analysis, we fit each line in the [{\sc O iii}]$\lambda\lambda4959,5007$ doublet with a coaddition of the outflow $\phi_{\rm \lambda o.f.}$ defined in Equation \ref{Equation:prof} and a Gaussian core $\phi_{\rm \lambda Gauss}$ to account for random motions of nebular gas not contained in the outflow, i.e.,
\begin{equation}
\begin{split}
    F_{\lambda} = F_0 \left[f\phi_{\rm \lambda Gauss} + (1-f)\phi_{\rm \lambda o.f.}\right]+C_\lambda
\end{split}
\end{equation}
where $f$ is the fraction of total flux $F_0$ contained in the Gaussian component of the line and $C_\lambda$ is the continuum flux. Fitting the [{\sc O iii}] doublet lines provides three advantages: (i) both are well-detected; (ii) both are emitted by the same ion species with identical excitation mechanisms, meaning the physical parameters for $\phi_{\rm \lambda Gauss}$ and $\phi_{\rm \lambda o.f.}$ are the same; and (iii) the $F_0$ and $\lambda_0$ of each line are related to those of the other by well-known atomic physics. Therefore, including both lines doubles the amount of data used the fit without adding additional variables or substantial uncertainty, thereby providing more robust constraints on the parameters.

\begin{figure}
    \centering
    \includegraphics[width=\linewidth]{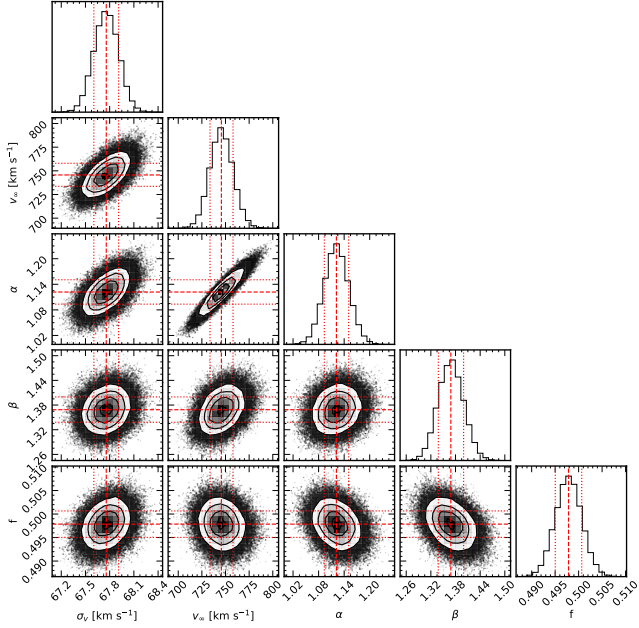}
    \caption{Posterior of the dynamically-relevant profile parameters for the [\ion{O}{iii}]$\lambda\lambda4959,5007$ doublet. %Flux is in units of $\rm 10^{-17}~erg~s^{-1}~cm^{-2}$, continuum $\rm 10^{-17}~ erg~s^{-1}~cm^{-2}~\AA^{-1}$ 
    Velocities in units of $\rm km~s^{-1}$. Best-fit values are indicated by red dashed lines while 0.1587 and 0.8413 quantiles are shown as red dotted lines. Values relevant to the dynamics of the ionized gas are listed Table \ref{tab:mcmc_fits}.}
    \label{fig:corner}
\end{figure}

\begin{figure}
    \centering
    \includegraphics[width=\linewidth]{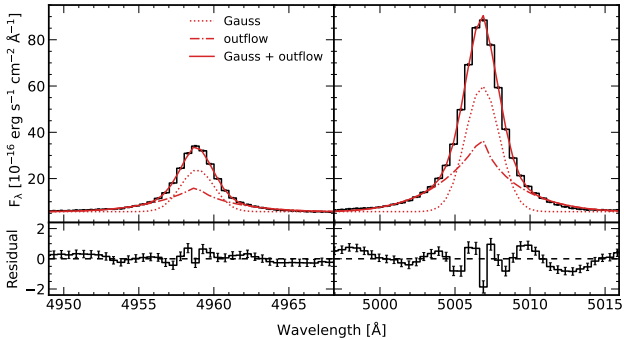}
    \caption{{\it Top}: Best-fit line profiles (red) overlaid on the [\ion{O}{iii}] doublet flux measured in the nucleus of Mrk 462 (black). Outflow component is the dash-dotted line. Gaussian component is the dotted line. Total is the solid line. {\it Bottom}: Residual flux obtained by taking the difference between the total fit profile and the observed flux. Error bars in all panels indicate the $3\sigma$ uncertainty in flux density in each pixel.}
    \label{fig:linefit}
\end{figure}

We sample the posterior on parameters $\lambda_0$, $\sigma_v$, $v_\infty$, $\alpha$, $\beta$, $f$, and $C_\lambda$ using the python MCMC package {\sc emcee} \citep{2013PASP..125..306F}. For robustness, we use 32 walkers, assume a burn-in of $10^3$ steps, and a take a subsequent $10^4$ steps. We assume a uniform prior with the requirements that fluxes be positive, $\alpha$ and $\beta$ remain in the possible range of values { (0 to 9 and 0 to 4, respectively, following the literature evaluations discussed in \S\ref{sec:profiles}, cf. \citealt{2016MNRAS.463..541C})}, and the Gaussian velocity width $\sigma_v$ not exceed $v_\infty$. Results of this sampling are listed in Table \ref{tab:mcmc_fits} with a posterior map shown in Figure \ref{fig:corner}. We show the corresponding best-fit profile in Figure \ref{fig:linefit}.

\begin{table}
    \centering
    \caption{Quantiles of the posterior of parameters for the combined Gaussian and outflow profiles for the [\ion{O}{iii}] profile observed in Mrk 462.}
    \label{tab:mcmc_fits}
    \begin{tabular}{l SSS}
    & \multicolumn{3}{c}{quantile} \\
    & 0.16 & 0.5 & 0.84 \\
    \hline
    $\sigma_v$ [$\rm km~s^{-1}$] & 67.604 & 67.758 & 67.911\\
    $v_\infty$ [$\rm km~s^{-1}$] & 733.629 & 745.635 & 757.960 \\
    $\alpha$ & 1.093 & 1.122 & 1.151 \\
    $\beta$ & 1.338 & 1.369 & 1.400 \\
    $f$ & 0.495 & 0.498 & 0.501 \\    
         %& [\ion{O}{iii}] & [\ion{N}{ii}] & H$\alpha$ \\
         %\hline
         %$\lambda_0$ & 5006.8\\
         %$\sigma_v$ ($\rm km~s^{-1}$) & $67.75\pm0.15$ & $48.87\pm$ & $50.07\pm$ \\
         %$v_\infty$ ($\rm km~s^{-1}$) & $745.38\pm12.15$ & $467.38\pm6.60$ & $667.04\pm6.90$\\
         %$\alpha$ & $1.12\pm0.03$ & $0.806\pm0.031$  & $1.321\pm0.020$ \\
         %$\beta$ &  $1.36\pm0.03$ & $1.230\pm0.035$ & $2.354\pm0.035$ \\
         %$f$ & $0.498\pm0.003$ & $0.5319\pm0.004$ & $0.499\pm0.002$
    \end{tabular}
\end{table}

The constraints on each of the parameters are quite good, with uncertainties of only a few percent. Moreover, the posterior samples appear quite normal and, by extension, multivariate normal. While most parameters exhibit a mild covariance, the $\alpha$ and $v_\infty$ posteriors are strongly positively correlated. Physically, this correlation may indicate that the density profile affects the terminal velocity of the outflow. The implication of such a relationship is that the acceleration of clouds in the outflow persists well beyond the initial launch radius.

{As evident in Figure \ref{fig:linefit}, the profile reproduces the asymmetric wings of the [\ion{O}{iii}] doublet with great accuracy. The best-fit profile recovers the observed doublet flux with a median residual of $2.47$\%. Median uncertainties in the flux are $1.37$\%, meaning that, characteristically, the profile flux is within $2\sigma$ of that observed. Such excellent agreement demonstrates that our flexible outflow profile model successfully describes the observed profile without making assumptions about the underlying mechanisms.}

\section{Interpreting the Outflow Profile}\label{sec:disc}

Having successfully fitted the Mrk 462 [\ion{O}{iii}] doublet with our outflow-motivated line profile, we now have the opportunity to evaluate the physical nature of the outflow.

First and foremost, we need to demonstrate how the best-fit parameters translate into properties of the outflow, namely the mass outflow rate and momentum and energy injection rates. Moreover, we are concerned with how the derived properties for Mrk 462 compare to properties of outflows observed in other galaxies. Placing Mrk 462 into a more general context of galactic outflows in this manner is critical for two reasons: (i) demonstrating that our novel analysis of the [\ion{O}{iii}] emission line profile is capable of providing insight into outflow properties and (ii) determining whether the outflow Mrk 462 scales appropriately with those observed in other galaxies. For these purposes, we draw on results in the literature for large samples of galactic outflows as measured via \ion{Na}{i} D \citep[][]{2005ApJS..160...87R,2015ApJ...809..147H} and UV absorption lines \citep[][]{2017MNRAS.469.4831C,2022ApJ...933..222X}.

{ Because these measurements span various ISM phases, comparing them to our own results bears justification. \citet{2005ApJS..160...87R} demonstrate in their Figure 17 that the \ion{Na}{i} D absorption line profiles are consistent with the blue-shifted wings of [\ion{N}{ii}] and [\ion{O}{iii}] emission lines in the same galaxies, indicating that both the neutral and ionized species trace galactic winds with similar outflow velocities. Similarly, \citet{2015ApJ...809..147H} find comparable profile shapes among various UV absorption lines arising from ions having a wide range of ionization potentials (8.151 to 33.492 eV). Previous studies have compared outflows measured by \ion{Na}{i} D and UV absorption lines from ion species of different ionization potentials \citep[e.g.,][]{2022ApJ...933..222X}, finding relatively consistent agreement between galaxy properties like stellar mass and star formation rate and outflow properties like velocity and momentum injection rate. Therefore, we can confidently compare results from measurements of various outflow phases to those from the ionized outflow traced by [\ion{O}{iii}] in Mrk 462.}

Secondly, we need to connect the best-fit velocity and density profiles to the underlying astrophysical processes. Specifically, what drives the outflow, and what are its effects on its host galaxy? To answer these questions, we draw on physically-based velocity profiles derived by \citet{2005ApJ...618..569M} from the equation of motion for a variety of driving mechanisms in order to assess the implications of the observed line profile.

\subsection{Outflow Momentum and Energy\label{sec:outprops}}

To determine the outflow momentum and energy, we follow the formalism outlined in \citet{2002ApJ...570..588R,2005ApJS..160..115R}. First, we determine the mass outflow rate from the column density (found by integrating the density profile from $R_0$ to $r$) and outflow velocity such that
\begin{equation}\label{Equation:mdot}
    \dot{M} = \Omega R_0 \mu m_{\rm H} N_{\rm H} v_{out} = 4\pi R_0^2 \mu m_{\rm H} n_0 v_\infty \mathscr{R}(x;\alpha,\beta) % v_\infty \left(\frac{R_0}{r}\right)^{1-\alpha}\left(1-\frac{R_0}{r}\right)^\beta.
\end{equation}
for solid angle $\Omega=4\pi$ sr \citep[cf. Equations 1-3 in][]{2005ApJS..160..115R}.  
%assuming the wind is thick, spherical, and isotropic (solid angle $\Omega=4\pi$).
%and that the mass density is given by $\rho = \mu m_{\rm H} n(r)$. 
The function $\mathscr{R}$ expresses the radial dependence of $\dot{M}$ where
\begin{equation}
    \mathscr{R}(x;\alpha,\beta) = 
    \begin{cases}
    \ln(x) \left(1-\frac{1}{x}\right)^\beta & \alpha = 1 \\
    \frac{1}{1-\alpha} \left[\left(\frac{1}{x}\right)^{\alpha-1}-1\right] \left(1-\frac{1}{x}\right)^\beta&\alpha\neq 1
    \end{cases}
\end{equation}
from the velocity profile (Equation \ref{Equation:velprof}) and the integral of the density profile (Equation \ref{Equation:denprof}) with respect to $x=r/R_0$. Two key differences distinguish our expression for $\dot{M}$ from that given by \citet{2002ApJ...570..588R,2005ApJS..160..115R}: (i) we do not assume an isothermal potential ($\alpha=2$) and (ii) we do not assume a radially constant velocity profile. 

Previous studies of outflows have either assumed some value of $R_0$ \citep[often 1-5 kpc, e.g.,][]{2005ApJS..160...87R} or used the UV half-light radius (i.e., the ``starburst'' radius) to determine $R_0$ under the premise that this value is similar to the launch radius regardless of mechanism \citep[e.g.,][]{2015ApJ...809..147H,2022ApJ...933..222X}. For $R_0$, we consider the SDSS $u$ band Petrosian radius of 2.7$^{\prime\prime}$, which corresponds to $\sim0.54\rm~kpc$ for Mrk 462. This value is consistent with UV half-light radii measured for other galactic outflows \citep{2015ApJ...809..147H,2022ApJ...933..222X}, which suggests our assumed $R_0$ is appropriate. However, \citet{2015ApJ...809..147H} argue that $R_0$ must be scaled in order to properly derive outflow properties. At what radius do we evaluate $\dot{M}$? Both $x=1$\ (i.e., $r=R_0$) and $x\to\infty$ will yield meaningless mass outflow rates (see discussion in \S\ref{sec:profiles}). We therefore choose the value $x_{out}$ which maximizes the momentum density $n v$. For the best-fit values of $\alpha$ and $\beta$, we list results for $x_{out}$ and the corresponding outflow velocity $v_{out}$ in Table \ref{tab:outflow}. Interestingly, our value for $x_{out}$ is close to the factor of 2 assumed in previous studies \citep[e.g.,][]{2015ApJ...809..147H,2022ApJ...933..222X}; however, given the dependence of $x_{out}$ and the related momentum distribution on $\alpha$ and $\beta$, we caution against simply assuming a factor of 2. In Appendix \ref{pdx:xout}, we elaborate on the effects of $\alpha$ and $\beta$ on $x_{out}$.

We obtain the particle density $n_0$ from an emission measure argument. Following Equation \ref{Equation:emiscoef}, the emission measure $n_e n_p$ can be obtained using the emission coefficient and luminosity for [\ion{O}{iii}]$\lambda5007$ such that
\begin{equation}\label{Equation:dens_em}
    n_0^2 \approx n_e n_p = \frac{L_{\rm \lambda5007}(1-f)}{4\pi \epsilon R_{0}^2\Delta R\varepsilon_{\rm \lambda5007}} \frac{\rm H^+}{\rm O^{++}}
\end{equation}
where the ionic abundance $\rm O^{++}/ H^+$ scales $n_{\rm O^{++}}$ to $n_p\approx n_e$, $\Delta R\sim1\rm~pc$ is the characteristic cloud size \citep[see, e.g.,][]{1994ApJ...420..213K,2006ApJS..164..477N}, and $\epsilon\sim0.01$ is the fraction of the outflow volume populated by clouds \citep[i.e., the filling factor, e.g.][]{1991RPPh...54..579O}\footnote{To calculate $n_0$ from the expected value of the density, one scales $n_0$ by a factor of $(\alpha+2)^{1-1/\alpha}$. For the value of $\alpha$ observed in the Mrk 462 [\ion{O}{iii}] doublet, this factor is roughly unity.}. The ionic abundance ${\rm O^{++}}/{\rm H^+}$ can be obtained using the flux and emissivity ratios for [\ion{O}{iii}] and H$\beta$ such that
\begin{equation}
    \frac{\rm O^{++}}{\rm H^+} = \frac{ F_{\rm\lambda5007}}{F_{\rm H\beta}}\frac{\varepsilon_{\rm H\beta}}{\varepsilon_{\rm\lambda5007}}
\end{equation}
for some assumed electron temperature \citep[Equation 5.41][see also Equation 16 in \citealt{Peimbert_2017}]{source:osterbrock2006}. Using the flux ratio $F_{\rm[\ion{O}{iii}]\lambda5007}/F_{\rm H\beta}=2.92$ reported in \citet{2014AJ....148..136M} and $\varepsilon$ calculated using {\sc PyNeb} \citep{2015A&A...573A..42L} assuming $1.5\times10^4$ K for the electron temperature { (based on our estimate that the [\ion{O}{iii}]$\lambda4363$ auroral line flux is $\approx2$\% that of the $\lambda4959,5007$ doublet, which is consistent with analysis of the SDSS spectrum included in \citealt{2014AJ....148..136M})}, we obtain $\rm O^{++}/ H^+=3\times10^{-5}$. { Finally, we obtain a density of $n_0\approx 41\rm~cm^{-3}$}, which is similar to the densities of 20 to 30 $\rm cm^{-3}$ in outflows \citet{2018MNRAS.481.1690C} studied using photoionization models.
% \footnote{The [\ion{O}{iii}]$\lambda4363$ line is only weakly detected (S/N=2.3) and thus cannot be fit robustly like the [\ion{O}{iii}]$\lambda\lambda4959,5007$ doublet. Scaling the best-fit [\ion{O}{iii}]$\lambda5007$ profile to match the $\lambda4363$ line using variance-weighted least squares gives a flux $\approx2$\% that of $\lambda5007$, implying $T_e\sim1.5\times10^4$ K.}

With density and radius in hand, we can calculate the mass outflow rate $\dot{M}$. We find $\dot{M} = 179 \rm~M_\odot~yr^{-1}$ for Mrk 462. The momentum and energy injection rates are given by Equations 9 and 11, respectively, in \citet{2005ApJS..160..115R} where the former is
\begin{equation}\label{Equation:pdot}
    \dot{p} = \dot{M} v_{out}
\end{equation}
and the latter is
\begin{equation}\label{Equation:edot}
    \dot{E} = \dot{M}\left[\frac{1}{2} v_{out}^2 + \frac{3}{2} \sigma_v^2 \right].
\end{equation}
We assume (somewhat na\"ively) that the turbulent energy within the outflow is equivalent to the velocity dispersion of the Gaussian component fit to the profile in the previous section. Results for $\dot{p}$ and $\dot{E}$ are listed in Table \ref{tab:outflow}.

\begin{table}
    \caption{Outflow properties derived from best-fit outflow profile for $n_0=41\rm~cm^{-3}$ % $10^{2.53}$ measured from the [\ion{S}{ii}] doublet 
    and $R_0=0.539$ kpc from the {\it u}-band Petrosian radius. Quantiles determined by propagating the MCMC samples of the posterior on the outflow profile parameters.}
    \label{tab:outflow}
    \centering
    \begin{tabular}{l SSS}
    & \multicolumn{3}{c}{quantile} \\
        & 0.16 & 0.5 & 0.84 \\
        \hline
                                        $x_{out}$ & 2.183 &  2.220 & 2.258\\
                                        $v_{out}\rm~[km~s^{-1}]$ & 326.888 & 328.619 & 330.385 \\
%                                $\log N_{\rm H}\rm~[cm^{-2}]$ & 20.201 20.213 20.226 \\
                $\mathscr{R}(x_{out};\alpha,\beta)$ & 0.321 & 0.335 & 0.350 \\

 $\log~\dot{M}\rm~[M_\odot~yr^{-1}]$ & 2.739 & 2.752 & 2.766 \\
            $\log~\dot{p}\rm~[dyne]$ & 36.055 & 36.068 & 36.083 \\
              $\log~\dot{E}\rm~[erg~s^{-1}]$ & 43.322 & 43.337 & 43.352 \\
    \end{tabular}
\end{table}

\begin{figure}
    \centering
    \includegraphics[width=\columnwidth]{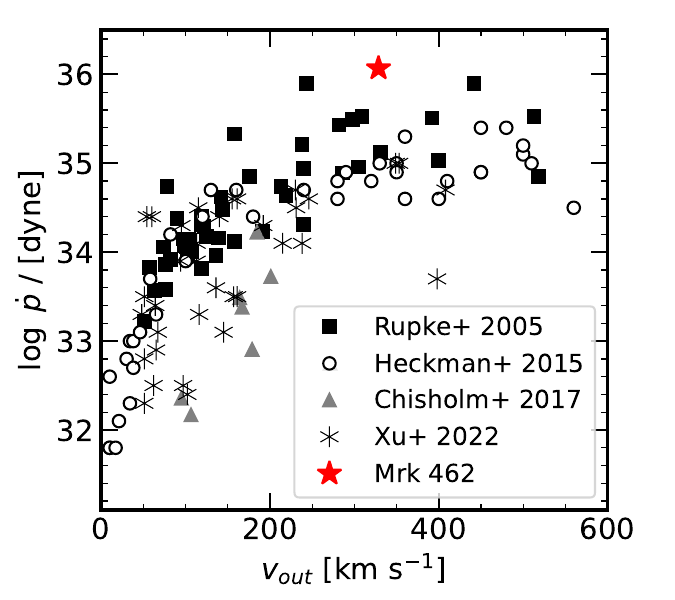}
    \caption{Momentum flux and velocity of the outflow in Mrk 462 (red star) compared to outflows observed in star-forming galaxies via absorption lines (squares \citealt{2005ApJS..160..115R}, open circles \citealt{2015ApJ...809..147H}, triangles \citealt{2017MNRAS.469.4831C}, asterisks \citealt{2022ApJ...933..222X}). Uncertainties obtained by propagating the MCMC samples of the best-fit parameters are smaller than the symbol for Mrk 462.}
    \label{fig:outflow}
\end{figure}

We compare the outflow properties in Table \ref{tab:outflow} to outflow properties in other galaxies in Figure \ref{fig:outflow}. As evident in Figure \ref{fig:outflow}, we find that the outflow in Mrk 462 has a momentum flux comparable to those of outflows observed in starburst galaxies via absorption lines. This result is compelling for two reasons. First, it validates our approach to deriving outflow properties from the observed line profile. Second, it places Mrk 462 into a rich context of starbursts, ULIRGs, and other star-forming galaxies. While we have not yet addressed what drives the outflow in Mrk 462, the outflow appears to be similar to those found in other galaxies. %The high outflow velocity might be due to a number of factors, including (i) AGN augmentation of the outflow by means of radiation pressure; (ii) systematic enhancement of the derived outflow due to the assumed spherical geometry implicit to our outflow model (essentially reducing the factor of $4\pi$ in Equation 13 to represent a conic opening angle, e.g., \citealt{2012ApJ...750...55S}); (iii) particularly high particle densities in the outflowing ionized gas in Mrk 462; and/or (iv) variations in the velocity or density gradient. We explore the effects of the assumed initial density and slopes of the outflow profiles on the distribution of derived momentum flux below.

%{ maybe also $\dot{M}/L_{H\alpha}$ as a proxy for mass loading and compare to adjusted results from Lochhaas+ 2021}

%{ also show relative to the circular velocity $v_{circ}\propto M_\star$, critical $\dot{p}$, and potential luminosity $m G M_\star\dot{M}/r$ for halo-to-stellar mass ratio $m\sim0.1$}

\begin{figure}
    \centering
    \includegraphics[width=\columnwidth]{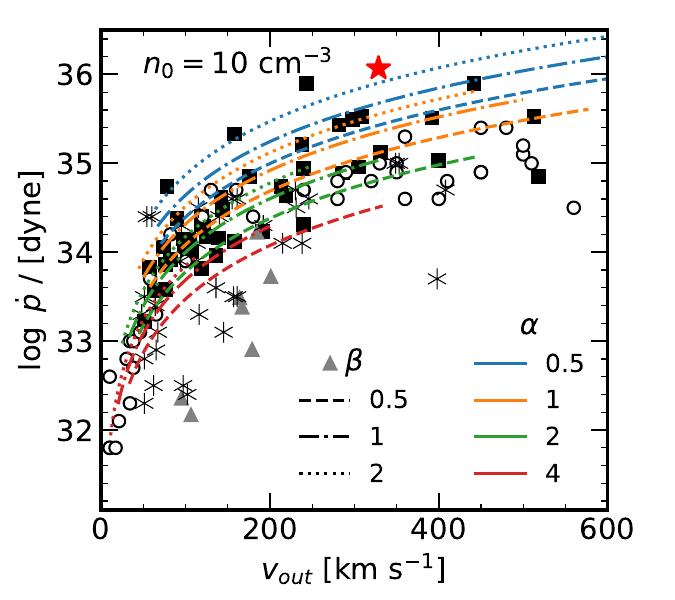}
    \caption{Effects of varying density profile slope $\alpha$ and velocity profile slope $\beta$ on the observed momentum flux and outflow velocity for values of $v_\infty$ ranging from 100 to 1,000 $\rm km~s^{-1}$. Symbols as in Figure \ref{fig:outflow}.}
    \label{fig:out_discuss}
\end{figure}

{The momentum fluxes and velocities of galaxy outflows form a sequence in Figure \ref{fig:outflow}, with momentum flux increasing with increasing outflow velocity. The outflow in Mrk 462 occupies the high-velocity end of this sequence where the momentum flux changes much more gradually with velocity than it does below $\sim200$ km s$^{-1}$. To determine if outflows with momentum fluxes and velocities similar to those of Mrk 462 can be characterized by similar density and velocity profiles, we predict sequences of $\dot{p}$ from Equations \ref{Equation:mdot}-\ref{Equation:pdot} over a range of $v_\infty$ for different values of $\alpha$ and $\beta$ assuming a density of $n_0=10$ cm$^{-3}$. We compare these predicted sequences of outflow properties to the observed distribution in Figure \ref{fig:out_discuss}. As $\alpha$ decreases and/or $\beta$ increases, the momentum flux and outflow velocity increase. This result suggests that the scatter in outflow $\dot{p}$ is real and arises from distinct astrophysical scenarios such as wind-driven fountains or radiation-driven shells. While density may also account for the observed scatter, the density and velocity profiles combined in Equations \ref{Equation:mdot}-\ref{Equation:pdot} readily describe the shape of the distribution of $\dot{p}$ over $v_{out}$. Thus, the sequence of outflow properties demonstrates a relationship between the underlying physics and the velocity and density profiles of the outflow.}

\begin{figure}
    \centering
    \includegraphics[width=\columnwidth]{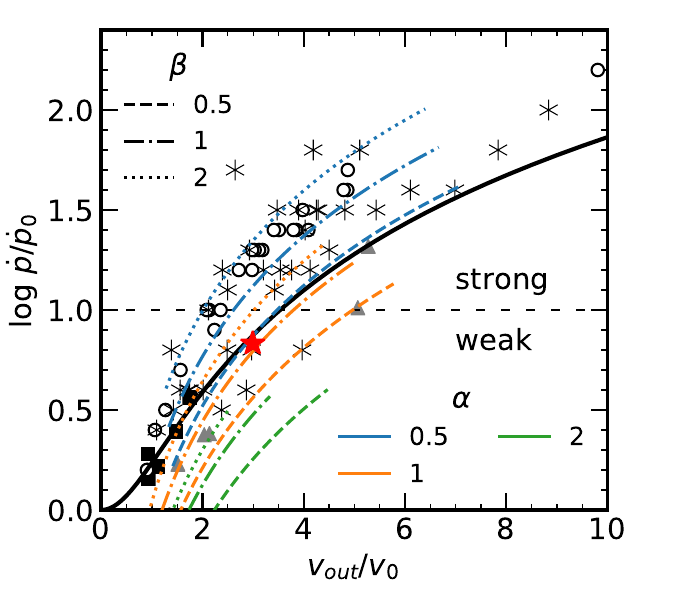}
    \caption{Momentum injection rate normalized to the critical momentum injection rate necessary to exceed the gravitational potential vs outflow velocity normalized to the rotational velocity of the galaxy inferred from the stellar mass. Thin dashed horizontal line is the demarcation between weak and strong outflows from \citet{2015ApJ...809..147H}. Solid line is the prediction for momentum-driven winds from \citet{2015ApJ...809..147H}. Predictions for various $\alpha$ and $\beta$ values correspond to $v_\infty/v_0\in[2,10]$. Symbols as in Figure \ref{fig:outflow} but including only galaxies with $\dot{p}>\dot{p}_0$.}
    \label{fig:out_norm}
\end{figure}

To understand these outflow properties in the context of the host galaxy, one conventionally expresses the outflow velocity and momentum injection rate relative to dynamical values \citep[see, e.g.,][]{2015ApJ...809..147H,2022ApJ...933..222X}. For the velocity, the fiducial dynamical value is the circular velocity $v_0$. We obtain the circular velocity for Mrk 462 using the stellar mass of $M_\star=5.5\times10^9$ M$_\odot$ reported by \citet{2014AJ....148..136M} and the empirical Tully-Fischer relation given by Equation 27 in \citet{2011MNRAS.417.2347R} such that
\begin{equation}
    \log v_0 = 2.141+0.278\left[\log(M_\star)-10.102\right]
\end{equation}
with an accuracy of 0.096 dex\footnote{Note that this relation is specific to the photometric mass-to-light ratios used in \citet{2014AJ....148..136M} to obtain the stellar mass of Mrk 462. That being said, results from this relation are within a few $\rm km~s^{-1}$ to the approach used in \citet{2015ApJ...809..147H,2022ApJ...933..222X}, well within the $\sim0.1$ dex uncertainty in both relations.}. Following \citet{2015ApJ...809..147H}, the critical injection rate of momentum to exceed the pull of gravity is given by
\begin{equation}
    \dot{p}_0 = 4\pi R_0^2 \mu m_{\rm H} n_0 v_0^2
\end{equation}
at the launch radius of the outflow. Combining the above expression with Equations \ref{Equation:mdot} and \ref{Equation:pdot}, we obtain a relative momentum injection rate of
\begin{equation}\label{Equation:pdotnorm}
    \frac{\dot{p}}{\dot{p}_0} = \mathscr{R}(x;\alpha,\beta)\left(\frac{v_\infty}{v_0}\right)^2\left(1-\frac{1}{x}\right)^\beta,
\end{equation}
which we evaluate at $x=x_{out}$. {We show predicted values of $\dot{p}/\dot{p}_0$ over a range of $v_\infty/v_0$ in Figure \ref{fig:out_norm}. Strong outflows require $\alpha\le1$ and/or $\beta\ge1$ with $v_{out}/v_0\ga4$. Mechanisms associated with low $\alpha$ and high $\beta$ may be more likely to drive strong outflows. This coupling between mechanism and outflow strength could even indicate that fountains, which are characterized by high $\alpha$, might not be able to produce outflows capable of affecting the host galaxy.}

From Figure \ref{fig:out_norm}, we arrive at several key realizations. Mrk 462 exhibits a relatively weak outflow, falling short of the \citet{2015ApJ...809..147H} requirement that the momentum injection rate exceed the critical value due to gravity by more than an order of magnitude. Furthermore, Mrk 462 is \emph{typical} in its outflow properties: many other galaxies have comparable outflow strengths and relative velocities. Moreover, comparison with the \citet{2015ApJ...809..147H} prediction for the wind equation of motion suggests that the momentum injection rate in Mrk 462 can be explained \emph{solely} by a momentum-driven outflow. We address the mechanism which provides this momentum below in \S\ref{sec:outmech}.

\begin{figure}
    \centering
    \includegraphics[width=\columnwidth]{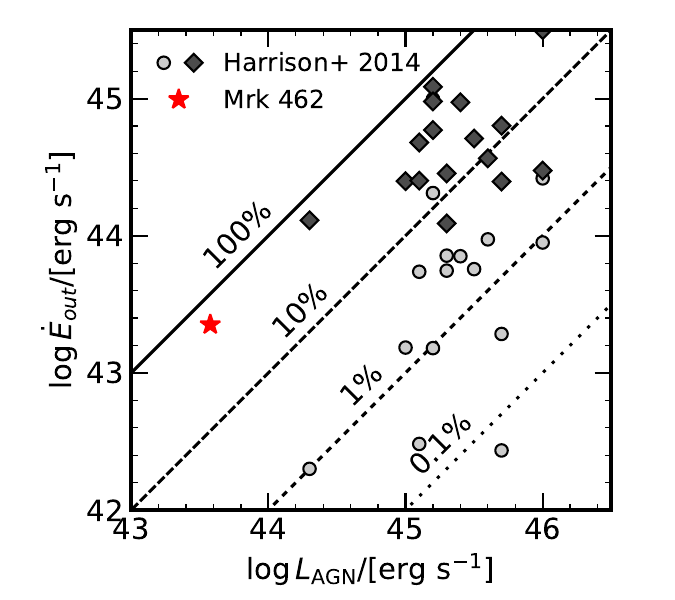}
    \caption{ Comparison of outflow mechanical energy $\dot{E}$ and bolometric AGN luminosity $L_{\rm AGN}$ for Mrk 462 (red star) and a sample of [\ion{O}{iii}]-traced AGN outflows from \citet{2014MNRAS.441.3306H} with $\dot{E}$ calculated using their methods 1 (light grey circles) and 2 (dark grey diamonds). We note that their method 2 is more comparable to ours than their method 1. Lines indicate constant values of outflow coupling efficiency.}
    \label{fig:edot_lagn}
\end{figure}

{ What remains to be learned from the outflow properties is the origin of the outflow. Mrk 462 exhibits both star-formation and AGN activity in its emission line spectrum, with young stars concentrated in a circumnuclear disk (Eleazer et al. in prep), meaning that nuclear star clusters and/or an accretion disk provide the momentum needed to produce the outflow. After correcting for extinction using the \citet{2016ApJ...828..107R} law and the reddening inferred from the Balmer decrement, the \emph{GALEX} \citep{2003SPIE.4854..336M} FUV flux suggests a star formation rate (SFR) of 0.25 $\rm M_\odot~yr^{-1}$ using the \citet{2013seg..book..419C} conversion for a 1-10 Myr burst. This SFR assumes that all of the FUV flux is produced by young stellar populations, i.e., with no AGN contribution. Even in this maximal case, the measured SFR is much lower than one would expect given published trends between SFR and mass outflow rate, outflow velocity, and momentum injection rates \citep[e.g.,][]{2015ApJ...809..147H,2022ApJ...933..222X}.

As discussed in \S\ref{sec:intro}, [\ion{O}{iii}] line profiles like that of Mrk 462 are found in many ``pure'' Seyferts. To determine if the outflow in Mrk 462 is consistent with AGN driving, we compare its energetics with those of [\ion{O}{iii}]-traced AGN outflows from \citet{2014MNRAS.441.3306H} in Figure \ref{fig:edot_lagn}, inferring the bolometric AGN luminostiy $L_{\rm AGN}=3.768\times10^{43}\rm~erg~s^{-1}$ for Mrk 462 from the flux of the narrow component of the [\ion{O}{iii}] line assuming the \citet{2004ApJ...613..109H} bolometric correction. Given that the \citet{2014MNRAS.441.3306H} method 2 is more comparable to our approach (i.e., it does not depend on the gas mass), this comparison suggests Mrk 462 is undergoing typical AGN-driven feedback close to 100\% outflow coupling efficiency (i.e., the AGN luminosity is closely linked to the outflow properties). We conclude, then, that the outflow is AGN-driven.}

\subsection{Outflow Driving Mechanisms\label{sec:outmech}}

A first look from Figure \ref{fig:out_norm} indicates that steep velocity and density gradients ($\beta\le1$ and $\alpha\ge2$, respectively) produce weaker outflows than shallow gradients. To understand the nature of strong vs weak outflows, both in the context of Mrk 462 and more generally, we want to know what underlying mechanism(s) and physical conditions produce or otherwise facilitate the observed profiles. The \citet{2015ApJ...809..147H} criterion suggests that mass-conserving or even mass-sweeping winds ($\alpha\le2$) produce the strongest outflows, especially if the velocity gradient is shallow (i.e., acceleration continues out to larger galactic radii).

\begin{figure}
    \centering
    \includegraphics[width=\columnwidth]{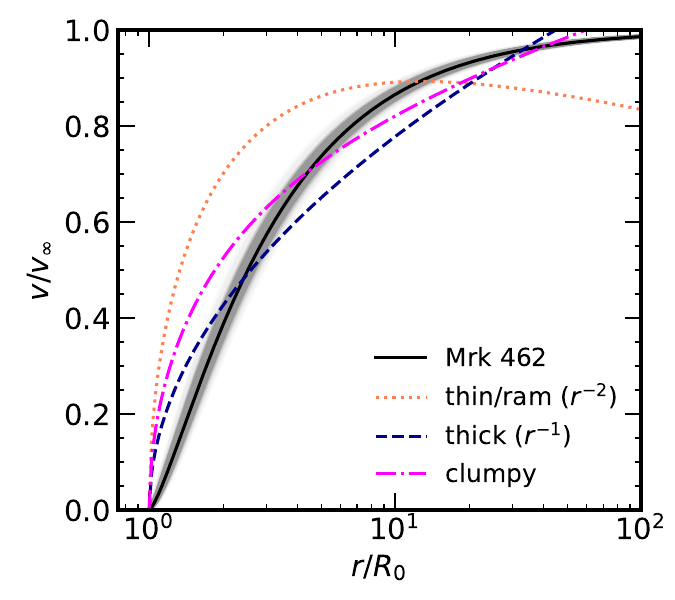}
    \caption{Velocity profile inferred from the [\ion{O}{iii}] line in Mrk 462 (solid black) compared to optically thick radiation pressure (dashed dark blue) and optically thin radiation or ram pressure (dotted orange) velocity profiles from \citet{2005ApJ...618..569M}, along with the profile expected in a dusty/clumpy radiation pressure scenario (dash-dotted magenta). Grey shaded region indicates the $3\sigma$ confidences on the shape of the profile.}
    \label{fig:velprof_mods}
\end{figure}

Now, we consider the possible mechanism(s) behind the outflow. \citet{2005ApJ...618..569M} define velocity profiles for several momentum-driven outflow mechanisms: optically thin and thick radiation pressure and ram pressure. Ram pressure and optically thin radiation pressure are formally similar, with acceleration $\propto r^{-2}$ and a velocity profile of
\begin{equation}\label{Equation:velthin}
    \left(\frac{v}{2v_0}\right)^2 = \frac{L}{L_{E}} \left[1-\frac{1}{x}\right] - \ln(x)
\end{equation}
where $L_{E}\propto v_0^2$ is the Eddington luminosity at the launch radius and $L$ is the luminosity of the driving mechanism\footnote{In the case of ram pressure, $L/L_E$ is equivalent to the $V_c^2/\sigma^2$ term in Equation A2 of \citet{2005ApJ...618..569M}.}. If, instead, the radiation pressure is optically thick, then the velocity profile becomes
\begin{equation}\label{Equation:velthick}
    \left(\frac{v}{2v_0}\right)^2 = \left[\frac{L}{L_{M}}-1\right]\ln(x)
\end{equation}
where $L_{M}\propto \sigma_v^4$ is the lower limit on luminosity for a momentum driven wind. We consider a third scenario which combines the optically thick and optically thin scenarios in the equation of motion. Since the thin and thick radiation pressure terms are separate, the resulting velocity profile is
\begin{equation}\label{Equation:velclump}
    \left(\frac{v}{2v_0}\right)^2 = \frac{L}{L_{E}} \left[1-\frac{1}{x}\right] + \left[\frac{L}{L_{M}}-1\right]\ln(x).
\end{equation}
This third scenario corresponds to dusty radiation pressure driving, which is optically thick in the UV and optically thin in the infrared \citep[cf. formalism in][]{2015MNRAS.451...93I,2018MNRAS.476..512I,2021MNRAS.502.3638I}. One might alternatively interpret the above expression as an optically thin outflow populated by optically thick clouds (i.e., a ``clumpy'' outflow)\footnote{As discussed in \citet{2005ApJ...618..569M}, the ``optically thin'' model for ram pressure is by definition a clumpy outflow with clouds entrained in a hot wind.}. Given the complexities of the ISM, the third scenario is likely a combination of both clumpiness and dusty radiation pressure. { A substantial fraction of dust (in excess of 10\% of the pre-existing dust) is known to survive in shocked environments such as wind-blown bubbles \citep[e.g.,][]{2010ApJ...713..592E}, AGN jets \citep[e.g.,][]{2001MNRAS.328..848V}, and SNe remnants \citep[e.g.,][]{2020ApJ...902..135S}, indicating that even under extreme conditions, a dusty radiatively-driven outflow is a plausible scenario to consider.} We determine $L/L_{E}$ and $L/L_{M}$ for the best-fit $\beta$ velocity profile using linear least squares and list the results in Table \ref{tab:mechs}. We show best-fit model profiles normalized to $v_\infty$ in Figure \ref{fig:velprof_mods}.

\begin{table}
    \centering
    \caption{Parameters for different outflow driving mechanisms determined by least squares fitting of model profiles to the observed velocity profile.}
    \label{tab:mechs}
    \begin{tabular}{c c c}
        model &  $L/L_E$ & $L/L_M$ \\
        \hline
        thin / ram & 12.88 & -- \\
        thick & -- & ~5.07 \\
        dusty / clumpy & ~3.75 & ~2.95 \\
    \end{tabular}
\end{table}

From the fit results shown in Figure \ref{fig:velprof_mods}, the radiatively-driven optically thick and clumpy outflow models provide much better descriptions of the observed profile than that of the optically thin / ram pressure. As such, we conclude that the outflow in Mrk 462, for which
$\beta = 1.369$, is radiatively driven.%we interpret the observed $\beta=1.36$ profile in Mrk 462 as a radiatively driven outflow. %The $r^{-2}$ mechanisms require a far steeper velocity profile where acceleration occurs more rapidly \cite[cf. $\beta\la0.5$ result in][]{2016MNRAS.463..541C}.

In context of Figure \ref{fig:out_norm}, the association of higher $\beta$ with optically thick or dusty radiation pressure suggests that radiatively driven outflows are relatively strong at fixed $\alpha$. Given that dust is the primary source of the optical depth and that dust opacities are orders of magnitude higher in the UV than at longer wavelengths \citep[e.g.,][]{2003ARA&A..41..241D}, the optically thick outflow almost certainly must be driven by UV photons. Thus, the role of UV luminosity as a feedback mechanism in Mrk 462, and galaxies hosting star formation and/or AGN in general, is pivotal. Effects due to these outflows will be particularly pronounced in dwarf galaxies where the gravitational potential is weak (and therefore $v_0$ is lower), resulting in much stronger feedback (recall $\dot{p}/\dot{p}_0\propto v_0^{-2}$ and $L/L_M\propto \sigma_v^{-4}$).

Which of the thick shell or dusty/clumpy geometries best describes the radiative outflow? Qualitatively, we favor the dusty/clumpy scenario simply because the uniform shell is highly idealized and oversimplifies the complexities of the ISM. However, the fits to the velocity profile for both uniformly optically thick and dusty radiation pressure scenarios shown in Figure \ref{fig:velprof_mods} seem comparably accurate. An initial quantitative assessment of the reduced sum squared residuals, $\chi^2_{\nu}$, shows that the dusty model outperforms the thick model by a factor of 2. To determine whether we can reject the simpler optically thick model on this basis, we calculate the log likelihood ratio $\lambda_r=\Delta\chi_{\nu}^2$. For sufficiently large data sets, $\lambda_r$ is drawn from a $\chi^2$ distribution with degrees of freedom equal to the difference in the number of parameters. Since we can make $x$ arbitrarily large, the survival function for the $\chi^2$ distribution approximates the probability that the two sets of residuals are similar, i.e., the probability that the uniform optically thick shell (the null hypothesis) is a sufficient description of the velocity profile when compared to the dusty shell (the alternate hypothesis). Reduced residuals from our fits give $\lambda_r=0.788$, which corresponds to a $p$-value of $0.376$. While our statistical assessment cannot rule out the null hypothesis that the optically thick mechanism alone drives the outflow, the dusty scenario more closely describes the velocity profile in Mrk 462. Thus, we favor a picture in which the outflow is dusty and radiatively driven.

The best fit parameters for the dusty outflow also provide insight into the relative luminosity of the driving mechanism (see Table \ref{tab:mechs}). For the optically thick component, we find $L/L_M\approx3$. When $L>L_M$, a galaxy experiences winds capable of removing gas, which can regulate star formation and/or black hole growth \citep[][]{2005ApJ...618..569M}. As a result, gas-fueled activity will likely be episodic. With $L/L_E\approx4$, the optically thin component of the outflow facilitates acceleration within a few $R_0$ of $x_{out}$ and contributes to the velocity profile for nearly an order of magnitude in radius \citep[see prediction that $L/L_E\sim3$ drives an outflow to at least $10R_0$ in][]{2005ApJ...618..569M} before optically thick radiation pressure becomes the sole source of momentum deposition.

{ While the velocity field can inform the mechanisms involved in driving the outflow, the best-fit driving luminosity $L$ can indicate more about the source of the outflow. To begin, the Eddington luminosity $L_E$ is given by
\begin{equation}
L_{E}=\frac{4\pi GM_\star c}{\kappa}.
\end{equation}
Because the AGN SED is typically most luminous in the FUV, we assume $\kappa\approx10^5\rm~cm^{2}~g^{-1}$, which is characteristic of dust opacity at $1000$ \AA\ \citep[e.g.,][]{2003ARA&A..41..241D}. We obtain an implied driving luminosity of $L=3.75L_{E}\approx10^{43}\rm~erg~s^{-1}$. \citet{2005ApJ...618..569M} give the critical luminosity $L_M$ as
\begin{equation}
    L_M = \frac{4f_gc\sigma_0^4}{G}
\end{equation}
where $f_g\sim0.001$ is the gas mass fraction estimated from the H$\beta$ flux following \citet{2014MNRAS.441.3306H} and the stellar mass from \citet{2014AJ....148..136M}. Assuming $\sigma_0$ is well-represented by the velocity dispersion determined from the narrow component of the [\ion{O}{iii}] line, we obtain a driving luminosity $L=2.95L_M\approx1\times10^{43}\rm~erg~s^{-1}$. Following the $L = 4.3\times10^{41}\rm SFR~erg~s^{-1}$ scaling relation in \citet{2022ApJ...933..222X}, the SFR in Mrk 462 obtained in \S\ref{sec:outprops} implies a luminosity of $10^{41}\rm~erg~s^{-1}$, which cannot account for either of the driving luminosities. However, the driving luminosity is sufficiently below the estimated $L_{\rm AGN}$ by about 0.5 dex, suggesting that AGN can account for the luminosities associated with the velocity profile.}

Our examination of both the relative properties and velocity profile of the outflow in Mrk 462 indicate several key aspects of the underlying mechanisms. First, the outflow is driven by radiation pressure imparting momentum to clouds. Second, dust likely provides the opacity for the radiation pressure, having both optically thin and optically thick components. The dusty clouds which constitute the outflow are more likely distributed in clumps than in a uniform shell. Third, the luminosity necessary to drive the dusty outflow is $3-4$ times the critical values, which can regulate gas-fueled activity in a galaxy. { Finally, the properties and velocity profile require source luminosities which stellar populations alone cannot produce, suggesting that the central AGN must provide the momentum needed to drive the outflow. Together, these characteristics indicate that Mrk 462 is producing a self-regulating, albeit relatively weak, radiatively-driven dusty outflow powered by the AGN.}

\section{Conclusion and Future Work}

Motivated by the asymmetric, broad forbidden nebular lines in Mrk 462, we have presented a physically motivated line profile based on the traditional outflow interpretation of similar profiles observed in other galaxies. We have demonstrated how underlying velocity and density profiles combine to yield the observed line profile, establishing a formalism which is flexible and readily applicable to other nebular lines. After application of this line profile to the [\ion{O}{iii}]$\lambda\lambda4959,5007$ doublet in Mrk 462, we illustrate how properties of the outflow can be derived, including characteristic momentum and energy injection rates. We also draw on previous model velocity profiles to interpret the best-fit velocity profile. From this comparison, we conclude that the outflow in Mrk 462 is relatively weak (although still quite strong), radiatively driven, and dusty/clumpy.

While we have provided a foundation for the analysis of outflows traced by asymmetric nebular emission lines, our results are merely that: foundational. Adjustments to our model to include additional effects such as dust may provide improvements to the line profile and results. Future work will also include application of our methods to the study of outflows in various contexts, specifically their role in galaxy evolution and the escape of ionizing photons.

Dust is likely entrained in the outflow, as evidenced by the radiative driving generated by the UV photon flux. The presence of dust could exacerbate the asymmetry in the outflow profile by preferentially extinguishing the emission from material moving away from the observer along the line of sight, thereby diminishing the red wing of the line profile relative to the blue \citep[cf.][]{1981ApJ...247..403H,1991ApJS...75..383V}. Such effects should be explored in detail in future work to determine if they are significant and, if so, how the inclusion of dust extinction within the outflow might affect the emergent line profile and derived outflow properties.

A velocity offset may exist between the Gaussian core and outflow profiles, too. Such an offset might arise from a rest velocity of the outflow launch which is not spatially coincident with the narrow Gaussian component. That is, the outflow may have a different rest velocity at the launch radius than the dynamically quiescent photoionized clouds in star-forming regions or the AGN NLR. Such a velocity offset might explain the red-winged outflow profiles observed in some galaxies \citep[e.g.][]{2021MNRAS.507...74R}.

One of the key applications of our methods will be the role of feedback in the escape of Lyman continuum photons. Previous studies of Lyman continuum emitters (LCEs) and Green Peas \citep[e.g.][respectively]{2017A&A...605A..67C,2017ApJ...851L...9J} found \emph{reduced} outflow velocities, indicative of suppressed superwinds. Indeed, the well-studied LCE Haro 11 has $v_{out}\approx160\rm~km~s^{-1}$ \citep{2015ApJ...809..147H}, less than half that of Mrk 462. However, the extreme [\ion{O}{iii}] wings in Mrk 71, a Green Pea and a LCE candidate, exhibit $v_\infty$ in excess of 3,000 $\rm km~s^{-1}$ \citep{2021ApJ...920L..46K}. The well-studied $z=2.37$ Sunburst Arc, a known LCE \citep{2019Sci...366..738R}, exhibits characteristic and terminal outflow velocities ($v_{out}=327~\rm km~s^{-1}$ and $v_\infty=750~\rm km~s^{-1}$, respectively) in its LyC-leaking regions nearly identical to those we found for Mrk 462 \citep{2022arXiv221011575M}. Similar high velocities have been found in [\ion{O}{iii}] profiles of local ($z\sim0.3$) known LCEs (Rodriguez et al. in prep, Komarova et al. in prep). The confounding results, as well as the ultimate cosmological implications for \emph{how} galaxies reionized the universe, should be addressed through use of our novel analysis methods. Doing so will determine whether Lyman continuum escape is facilitated by feedback and outflows, to what degree, and which physical mechanisms are involved.

Another important question to address will be the roles of AGN and star-formation feedback in galaxy evolution. Characterizing the spatial distribution of UV and IR fluxes in Mrk 462 and other galaxies with outflows will be key to understanding the nature of dusty radiative feedback. Analysis of outflow line properties in both AGN and starbursts will be key, particularly when both are present, to assess whether and how radiative outflows affect star formation and black hole growth. Investigating the evolution of these outflow properties over cosmic time will also be critical for assessing when outflows are at their strongest and what role they play in determining the stellar mass of their host galaxies.

\section*{Software}

{{\sc\ astropy\ } \citep{astropy:2013,astropy:2018}, {\sc corner\ } \citep{corner}, {\sc\ emcee\ } \citep{2013PASP..125..306F}, {\sc\ matplotlib\ } \citep{matplotlib}, {\sc\ numpy\ } \citep{numpy}, {\sc\ pyneb\ } \citep{2015A&A...573A..42L}, {\sc\ scipy\ } \citep{scipy}, {\sc\ sympy} \citep{sympy}}

\section*{Acknowledgements}

We thank the anonymous referee for feedback which improved the clarity of this paper.

\section*{Data Availability}

The calibrated data underlying this article were provided by Eleazer et al (in prep) by permission. The raw data are publicly available through the Gemini data archive at \url{https://archive.gemini.edu/searchform}.% Once Eleazer et al (in prep) have published the full set of observations, the data will be shared on reasonable request to the corresponding author with permission of Eleazer et al (in prep).

%%%%%%%%%%%%%%%%%%%% REFERENCES %%%%%%%%%%%%%%%%%%

% The best way to enter references is to use BibTeX:

\bibliographystyle{mnras}
\bibliography{bib} % if your bibtex file is called example.bib

% Alternatively you could enter them by hand, like this:
% This method is tedious and prone to error if you have lots of references
%\begin{thebibliography}{99}
%\bibitem[\protect\citeauthoryear{Author}{2012}]{Author2012}
%Author A.~N., 2013, Journal of Improbable Astronomy, 1, 1
%\bibitem[\protect\citeauthoryear{Others}{2013}]{Others2013}
%Others S., 2012, Journal of Interesting Stuff, 17, 198
%\end{thebibliography}

%%%%%%%%%%%%%%%%%%%%%%%%%%%%%%%%%%%%%%%%%%%%%%%%%%

%%%%%%%%%%%%%%%%% APPENDICES %%%%%%%%%%%%%%%%%%%%%
\onecolumn

\appendix

\section{Example Profile Implementation in {\sc\small python}}\label{pdx:code}

%$v_\infty$ is in $c$ units
\begin{lstlisting}[language=Python]
from numpy import zeros,array,sqrt,trapz
from scipy.optimize import brentq
from scipy.integrate import romberg
# function to calculate wind profile, assuming vinf in c units
def wind_profile(wave,wave0,vinf,alpha,beta):
    # observed velocity from wavelengths, in vinf/c units
    u = abs(wave-wave0)/wave0 /vinf
    # normalized radius from velocity
    x = lambda w: ( 1-w**(1/beta) )**-1
    # normalized density profile n/n0
    n = lambda w: ( 1-w**(1/beta) )**alpha
    # with no occultation, the minimum velocity is just the observed velocity
    wmin = u[u<1]
    # when occultation by source occurs, solve for minimum contributing velocity
    wmin[wave[u<1]>wave0] = array(list( map( lambda ui: \
        brentq(lambda wl: ui - wl/x(wl)*sqrt(x(wl)**2-1),ui,1), \
        u[(u<1)&(wave>wave0)]) ))
    # profile integrated at each velocity range
    phi = zeros(len(u))
    phi[u<1] = array(list( map( lambda wl: \ 
        romberg(lambda w: n(w)**2,wl,1), \ 
        wmin) ))
    # normalize so integral goes to 1 and units are wavelength^-1
    norm = trapz(phi,x=wave)
    return phi/norm
\end{lstlisting}

\section{Radius of Maximum Momentum Density}\label{pdx:xout}

\begin{figure}
    \centering
    \includegraphics[width=0.45\linewidth]{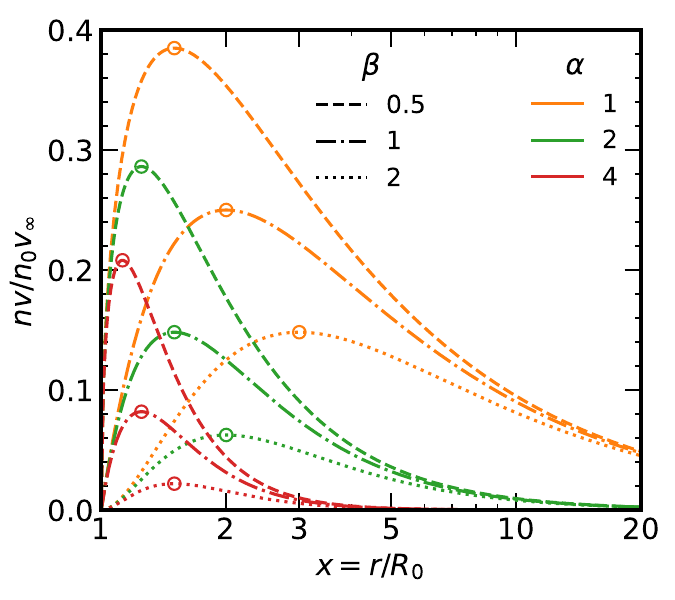}
    \includegraphics[width=0.49\linewidth]{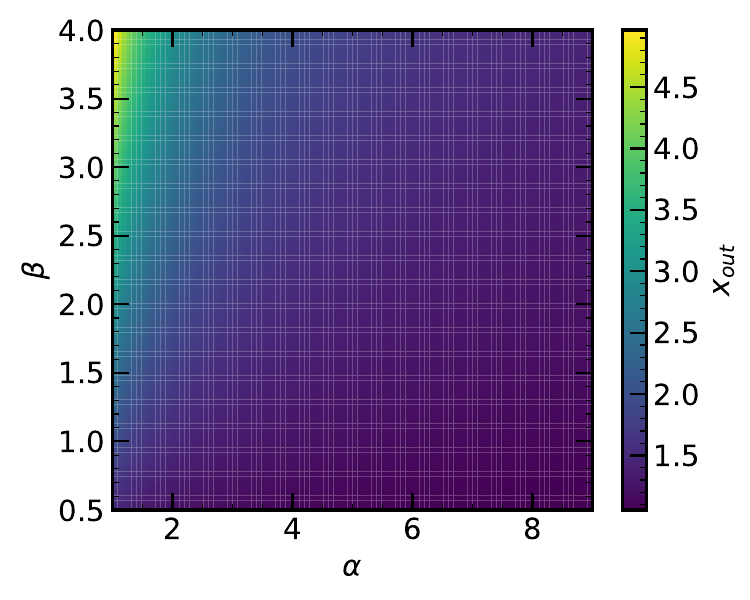}
    \caption{\emph{Left:} Momentum density profiles for different values of $\alpha$ and $\beta$. The location of maximum momentum density for each profile is indicated by an open circle. \emph{Right:} The outflow radius $x_{out}$ at maximum momentum density mapped over $\alpha$ and $\beta$.}
    \label{fig:xout}
\end{figure}

We show the momentum density as a function of radius in the left panel of Figure \ref{fig:xout}. Regardless of $\alpha$ and $\beta$, a maximum in $nv$ consistently occurs; however, the location of this maximum clearly varies with $\alpha$ and $\beta$. To obtain the radius $x_{out}$ at which the outflow momentum density is maximal, we calculate $x_{out}$ for the range of typical values of $\alpha$ and $\beta$ (see \S\ref{sec:profiles}) by root-finding. We visualize the effects of $\alpha$ and $\beta$ on the peak of the momentum distribution in the right panel of Figure \ref{fig:xout}. Values of $x_{out}$ increase with decreasing $\alpha$ and increasing $\beta$.

From a well-sampled grid of typical values of $\alpha$ and $\beta$, we find that $x_{out}\in(1,5]$ and that $x_{out}$ increases with $\beta$ and decreases with $\alpha$. We map values of $x_{out}$ over $\alpha$ and $\beta$ in the right panel of Figure \ref{fig:xout}. Typical values of $x_{out}$ are close (but not equal) to unity with an upper 95th percentile on $x_{out}$ of 2.80.

%%%%%%%%%%%%%%%%%%%%%%%%%%%%%%%%%%%%%%%%%%%%%%%%%%

% Don't change these lines
\bsp	% typesetting comment
\label{lastpage}
\end{document}